\documentclass[preprint,aps]{revtex4}
\usepackage{amssymb}

\begin{document}

\title{Asymptotically AdS Magnetic Branes in ($n+1$)-dimensional Dilaton Gravity}
\author{M. H. Dehghani$^{1,2}$\footnote{email address:
mhd@shirazu.ac.ir} and A. Bazrafshan$^1$}
\affiliation{$^1$ Physics Department and Biruni Observatory, College of Sciences, Shiraz University, Shiraz 71454, Iran\\
$^2$ Research Institute for Astrophysics and Astronomy of Maragha
(RIAAM), Maragha, Iran}

\begin{abstract}
We present a new class of asymptotically AdS magnetic solutions in ($n+1$)-dimensional dilaton gravity
in the presence of an appropriate combination of three Liouville-type potentials.
This class of solutions is asymptotically AdS in six and higher dimensions and yields a spacetime with longitudinal
magnetic field generated by a static brane. These solutions have no curvature singularity and no
horizons but have a conic geometry with a deficit angle. We find that the brane tension depends
on the dilaton field and approaches a constant as the coupling constant of dilaton field
goes to infinity. We generalize this class of solutions to the case of spinning magnetic solutions and find
that, when one or more rotation parameters are nonzero, the brane has a net electric charge which is
proportional to the magnitude of the rotation parameters. Finally, we use the counterterm method inspired
by AdS/CFT correspondence and compute the conserved quantities of these spacetimes. We found that the
conserved quantities do not depend on the dilaton field, which is evident from the fact that the
dilaton field vanishes on the boundary at infinity.
\end{abstract}

\maketitle

The AdS/CFT duality \cite{AdS/CFT} presents a powerful framework with which one may study
strongly coupled gauge theories. A large volume of
calculational evidence indicates that that a (relativistic) conformal field
theory (CFT) can be mapped to gravitational dynamics in an asymptotically
anti-de Sitter (AdS) spacetime of one larger dimension. One typical and
fascinating aspect of this duality is the property of holography, which
states that the amount of information contained in the boundary gauge theory
is the same as the one contained in the bulk string theory. This holographic
toolbox has since spread its usage to many branches of physics from
hydrodynamic \cite{Kov} and non-Fermi liquids \cite{Liu} to condensed matter systems \cite{Hart}. So, here we
are interested in asymptotically AdS horizonless solutions of dilaton
gravity.

Dilaton gravity can be thought as the low energy limit of string
theory. Indeed, in the low energy limit of string theory, one recovers
Einstein gravity along with a scalar dilaton field which is nonminimally
coupled to the gravity and other fields such as gauge fields \cite{String}.
The action of dilaton gravity also contains a Liouville-type potential,
which can be resulted by the breaking of spacetime supersymmetry in ten
dimensions. This Liouville-type potential also represents the higher
dimensional cosmological constant in non-critical string theories \cite{Lio}%
. In the presence of a Liouville-type potential, the solutions are neither
asymptotically flat nor AdS. This is due to the fact that maximally
symmetric spacetimes are not the solutions of field equations in the
presence of Liouville-type potential \cite{non-AdS}. Some attempts have been
made to explore various solutions of Einstein-Maxwell-dilaton (EMD) gravity
in the presence of Liouville-type potential \cite{Mann}. Also, the solutions
of dilaton gravity in the presence of two Liouville-type potentials have
been investigated by a number of authors both in the context of
Friedman-Robertson-Walker scalar field cosmology \cite{Cos} and
Einstein-Maxwell-dilaton (EMD) black holes \cite{2Lio}. Again in the presence of
two Liouville-type potentials, the solutions are neither asymptotically flat
nor (A)dS. Rotating black brane solutions with flat boundary in the
presence of one and two Liouville-type potentials have been also obtained in \cite{Rot}.

Besides investigating various aspects of black hole solutions, there has
been a lot of interest in studying the horizonless solutions in various
theories of gravity. Strong motivation for studying such kinds of solutions
comes from the fact that they may be interpreted as cosmic strings/branes.
Cosmic strings/branes are topological defects which are inevitably formed
during phase transitions in the early universe, and their subsequent
evolution and observational signatures must therefore be understood. The
string model of structure formation may help to resolve one of cosmological
mystery, the origin of cosmic magnetic fields \cite{Vil2}. There is strong
evidence from all numerical simulations for the scaling behavior of the long
string network during the radiation-dominated era. Apart from their possible
astrophysical roles, topological defects are fascinating objects in their
own right. These spacetime, which have flat hypersurface of $t=\mathrm{%
constant}$, $r=\mathrm{constant}$, have no curvature singularity, no
horizon, but they have a conic singularity. The 4-dimensional solutions of
this type in Einstein gravity with vanishing cosmological constant have been
considered in \cite{Levi}. Similar static solutions in the context of cosmic
string theory have been found in \cite{Vil}. All of these solutions \cite
{Levi,Vil} are horizonless and have a conical geometry; they are everywhere
flat except at the location of the line source. An extension to include the
electromagnetic field has also been done \cite{Muk}. Asymptotically AdS
spacetimes generated by static and spinning magnetic sources in three and
four-dimensional Einstein-Maxwell gravity with negative cosmological
constant have been investigated in \cite{Lem1}. These kinds of solutions in
dilaton gravity in the presence of one Liouville-type potential have been
constructed and their properties have been investigated \cite{Mag}.
Unfortunately, these solutions \cite{Mag} are neither asymptotically flat
nor (A)dS. The purpose of the present paper is to construct a new class of
static and spinning magnetic dilaton string solutions which produces a
longitudinal magnetic field in the background of anti-de Sitter spacetime.
Recently, some attempts have been made to construct solutions in dilaton
gravity which are asymptotically AdS with an appropriate combination of
three Liouville-type dilaton potentials \cite{Gao, 3Lio}. The four-dimensional
magnetic solution of EMD gravity in the presence of three Liouville-type potentials has been
introduced in \cite{Sheykhi-mag}. While this solution has not exact asymptotic AdS behavior, we like to
construct asymptotically AdS horizonless solutions in this theory in $(n+1)$ dimensions.

The outline of this paper is as follows. In the next section, we present a
brief review of the field equations of Einstein-Maxwell dilaton gravity. In
Sec \ref{charged}, we obtain the $(n+1)$-dimensional horizonless solutions
of the EMD gravity in the presence of three Liouville-type potentials and
investigate their properties. Section \ref{Spin} is devoted to the
horizonless spacetimes generated by spinning magnetic branes. We calculate
the conserved quantities of these spacetimes in Sec. \ref{Cons}. Finally, we
give some concluding remarks.

\section{Basic Field Equations \label{Field}}

The action of Einstein-Maxwell dilaton gravity in $(n+1)$ dimensions is
\begin{eqnarray}
I_{G} &=&-\frac{1}{16\pi }\int_{\mathcal{M}}d^{n+1}x\sqrt{-g}\left( \mathcal{%
R}{\ }-\frac{4}{n-1}(\nabla \Phi )^{2}-V(\Phi )-e^{-4\alpha \Phi
/(n-1)}F_{\mu \nu }F^{\mu \nu }\right)   \nonumber \\
&&-\frac{1}{8\pi }\int_{\partial \mathcal{M}}d^{n}x\sqrt{-h}K,  \label{Act}
\end{eqnarray}
where $\mathcal{R}$ is the Ricci scalar curvature, $\Phi $ is the dilaton
field, $\alpha $ is a constant determining the strength of coupling of the
scalar and electromagnetic fields, $F_{\mu \nu }=\partial _{\lbrack \mu
}A_{\nu ]}$ is the electromagnetic tensor field and $A_{\mu }$ is the vector
potential. The last term in Eq. (\ref{Act}) is the Gibbons-Hawking boundary
term which is chosen such that the variational principle is well-defined,
and $K$ is the trace of the extrinsic curvature $K^{ab}$ of any
boundary(ies) $\partial \mathcal{M}$ of the manifold $\mathcal{M}$, with
induced metric(s) $h_{ab}$. We consider the solutions of EMD theory in the
presence of the following potential
\begin{eqnarray}
{V}({\Phi }) &=&\frac{2\Lambda }{n(n-2+\alpha ^{2})^{2}}\{-\alpha ^{2}\left[
(n+1)^{2}-(n+1)\alpha ^{2}-6(n+1)+\alpha ^{2}+9\right] e^{{-4(n-2){\Phi }}%
/[(n-1)\alpha ]}  \nonumber  \label{V1} \\
&&+(n-2)^{2}(n-\alpha ^{2})e^{{4\alpha {\Phi }}/({n-1})}+4\alpha
^{2}(n-1)(n-2)e^{{-2{\Phi }(n-2-\alpha ^{2})}/[(n-1)\alpha ]}\},
\end{eqnarray}
where $\Lambda $ is the cosmological constant. It is clear that the
cosmological constant is coupled to the dilaton in a very nontrivial way.
This type of dilaton potential, which is a combination of three Liouville
potentials was introduced for the first time in \cite{Gao}. The motivation
for this dilaton potential is that the solutions in the presence of this
potential is asymptotically (A)dS in six and higher dimensions. Here, we are
interested in the asymptotically AdS solutions and therefore we define $%
\Lambda =-n(n-1)/2l^{2}$, where $l$ is the AdS radius of spacetime.

The field equations in $(n+1)$ dimensions are obtained by varying the action
(\ref{Act}) with respect to the gravitational field $g_{\mu \nu }$, the
dilaton field $\Phi $ and the gauge field $A_{\mu }$ :
\begin{equation}
\mathcal{R}_{\mu \nu }=\frac{4}{n-1}\left( \partial _{\mu }\Phi \partial
_{\nu }\Phi +\frac{1}{4}g_{\mu \nu }V(\Phi )\right) +2e^{{-4\alpha \Phi }/{%
(n-1)}}\left( F_{\mu \eta }F_{\nu }^{{\ }\eta }-\frac{1}{2(n-1)}g_{\mu \nu
}F_{\lambda \eta }F^{\lambda \eta }\right) ,  \label{FE1}
\end{equation}
\begin{equation}
\nabla ^{2}\Phi =\frac{n-1}{8}\frac{\partial V}{\partial \Phi }-\frac{\alpha
}{2}e^{-{4\alpha \Phi }/{(n-1})}F_{\lambda \eta }F^{\lambda \eta },
\label{FE2}
\end{equation}
\begin{equation}
\partial _{\mu }\left( \sqrt{-g}e^{{-4\alpha \Phi }/({n-1})}F^{\mu \nu
}\right) =0.  \label{FE3}
\end{equation}

\section{Static magnetic Branes \label{charged}}

In this section, we want to obtain the $(n+1)$-dimensional solutions of Eqs.
(\ref{FE1})-(\ref{FE3}) which produce longitudinal magnetic fields in the
Euclidean submanifold spans by $x^{i}$ coordinates ($i=1,...,n-2$). We will
work with the following ansatz

\begin{equation}
ds^{2}=-\frac{\rho ^{2}}{l^{2}}R^{2}(\rho )dt^{2}+\frac{d\rho ^{2}}{f(\rho
)g(\rho )}+l^{2}f(\rho )d\varphi ^{2}+\frac{\rho ^{2}}{l^{2}}R^{2}(\rho
)dX^{2}{,}  \label{Met1}
\end{equation}
where the constant $l$ have dimension of length and $dX^{2}={{%
\sum_{i=1}^{n-2}}}(dx^{i})^{2}$. Note that the coordinates $x^{i}$ ( $%
-\infty <x^{i}<\infty )$\ have the dimension of length, while the angular
coordinate $\varphi $ is dimensionless as usual and ranges in $0\leq \varphi
<2\pi $. The motivation for this metric gauge $[g_{tt}\varpropto -\rho ^{2}$
and $(g_{\rho \rho })^{-1}\varpropto g_{\varphi \varphi }]$ instead of the
usual Schwarzschild gauge $[(g_{\rho \rho })^{-1}\varpropto g_{tt}$ and $%
g_{\varphi \varphi }\varpropto \rho ^{2}]$ comes from the fact that we are
looking for a magnetic solution instead of an electric one. The Maxwell
equation (\ref{FE3}) can be integrated immediately to give

\begin{equation}
F_{\varphi \rho }=\frac{qle^{4\alpha \Phi /(n-1)}}{(\rho R)^{n-1}{g^{1/2}(\rho
)}},  \label{Fpr}
\end{equation}
where $q$ is the constant of integration. Using metric (\ref{Met1}) and the
Maxwell field (\ref{Fpr}), the field equations (\ref{FE1})-(\ref{FE3})
reduce to

\begin{eqnarray}
&&2\left( \frac{R^{\prime \prime }}{R}+3\frac{{R^{\prime }}^{2}}{R^{2}}+%
\frac{8R^{\prime }}{rR}+\frac{3}{r^{2}}\right) fg+\left( 2\,f^{\prime
}g+fg^{\prime }\right) \left( \frac{R^{\prime }}{R}+\frac{1}{r}\right) \frac{%
R^{\prime }}{R}  \nonumber \\
&&\hspace{4cm}+\frac{2}{n-1}\left\{ V+\,\frac{2{q}^{2}}{l^{2}(rR)^{2(\,n-2)}}%
e{^{4\alpha \Phi /(n-1)}}\right\} =0,  \nonumber \\
&&16\left( \frac{R^{\prime \prime }}{R}+\frac{2R^{\prime }}{rR}\right)
fg+2gf^{\prime \prime }+8(fg^{\prime }+gf^{\prime })\left( \frac{R^{\prime }%
}{R}+\frac{1}{r}\right) +g^{\prime }f^{\prime }  \nonumber \\
&&\hspace{4cm}+\frac{4}{n-1}\left\{ V+4fg{\Phi ^{\prime }}^{2}+\frac{2\left(
n-2\right) {q}^{2}}{l^{2}(rR)^{2\,(n-1)}}e{^{4\,\alpha \,\Phi /(n-1)}}%
\right\} =0,  \nonumber \\
&&2fg\Phi ^{\prime \prime }+\left\{ 8\left( \frac{R^{\prime }}{R}-\frac{1}{r}%
\right) \,fg+2\,f^{\prime }g-g^{\prime }f\,\right\} \Phi ^{\prime }+\frac{%
2\alpha {q}^{2}}{l^{2}(rR)^{2(\,n-1)}}e^{4\alpha \,\Phi /(n-1)}-\frac{n-1}{4}%
\frac{dV}{d\Phi }=0,  \nonumber \\
&&2gf^{\prime \prime }+\left\{ 8\left( \frac{R^{\prime }}{R}+\frac{1}{r}%
\right) \,g+g^{\prime }\right\} f^{\prime }+\frac{4}{n-1}\left\{ V+\frac{%
2\left( n-2\right) {q}^{2}}{l^{2}(rR)^{2(\,n-1)}}e^{4\alpha \,\Phi
/(n-1)}\right\} =0,  \label{EqsF}
\end{eqnarray}
where prime denotes the differentiation with respect to $\rho $. There are
four unknown functions $f(\rho )$, $g(\rho )$, $R(\rho )$ and $\Phi (\rho )$
in the above four field equations. One can show that the following solution
\begin{eqnarray}
f(\rho ) &=&\frac{\rho ^{2}}{l^{2}}\Gamma ^{\gamma }(\rho )-\left( \frac{c}{%
\rho }\right) ^{n-2}\Gamma ^{1-\gamma (n-2)}(\rho ),\nonumber \\
g(\rho ) &=&\Gamma ^{(n-3)\gamma }(\rho ), \nonumber\\
R(\rho ) &=&\Gamma ^{\gamma /2}(\rho ), \nonumber\\
\Phi (\rho ) &=&\frac{{(n-1)}{\alpha }}{{2}{(n-2+\alpha ^{2})}}\ln \Gamma(\rho ) \label{frho1}
\end{eqnarray}
satisfies all the components of field equations (\ref{EqsF}), where
\begin{eqnarray}
\Gamma (\rho ) &=&1+\left( \frac{b}{\rho }\right) ^{n-2}  \nonumber \\
\gamma  &=&\frac{2\alpha ^{2}}{\left( n-2\right) \left( n-2+\alpha
^{2}\right) }, \nonumber\\
q^{2} &=&\frac{\left( n-1\right) \left( n-2\right) ^{2}l^{2}}{2\left(
n-2+\alpha ^{2}\right) }(cb)^{n-2}, \label{frho2}
\end{eqnarray}
and $b$ and $c$\ are constants of integration, which are assumed to be
positive.

In order to study the general structure of these solutions, we first
consider the asymptotic behavior of these solutions. The metric (\ref{Met1}) at large $\rho$ can be written as
\begin{equation}
ds^{2}=-\frac{\rho ^{2}}{l^{2}}dt^{2}+\frac{d\rho ^{2}}{h_{\infty }(\rho )}%
+l^{2}f_{\infty }(\rho )d\varphi ^{2}+\frac{\rho ^{2}}{l^{2}}R^{2}(\rho
)dX^{2}{,}  \label{MetAsym}
\end{equation}
where
\begin{equation}
f_{\infty }(\rho )=\left\{
\begin{array}{cc}
h_{\infty }(\rho )=\frac{1}{l^{2}}\left\{ \left( \rho+\frac{\alpha ^{2}b}{1+\alpha ^{2}}\right)
^{2}-\frac{\alpha ^{2}b^{2}}{\left( 1+\alpha ^{2}\right) ^{2}}\right\}  & n=3
\\
h_{\infty }(\rho )-\frac{\alpha ^{2}b^{2}}{2+\alpha ^{2}}=\frac{\rho^{2}}{l^{2}}%
+\frac{\alpha ^{2}b^{2}}{2+\alpha ^{2}} & n=4 \\
\frac{\rho^{2}}{l^{2}} & n\geq 5
\end{array}
\right. .  \label{Asym}
\end{equation}
As one see from Eqs. (\ref{MetAsym}) and (\ref{Asym}), the solutions for $%
n\leq 4$ have not exact asymptotic AdS behavior, while they are
asymptotically AdS for $n\geq 5$. Indeed, the metric given by Eqs. (\ref{MetAsym})
and (\ref{Asym}) satisfy Einstein's equation with negative
cosmological constant only for  $n\geq 5$. In this paper we are interested in
asymptotically AdS solutions, and therefore we consider the solutions in six
and higher dimensions. Specially in Sec. \ref{Cons}, we use the counterterm
method inspired by AdS/CFT correspondence in order to calculate the
conserved quantities of the solutions.

Second, we look for curvature singularities. It is easy to show that the
Kretschmann scalar $R_{\mu \nu \lambda \kappa }R^{\mu \nu \lambda \kappa }$
diverges at $\rho =0$ and therefore one might think that there is a
curvature singularity located at $\rho =0$. However, as we will see below,
the spacetime will never achieve $\rho =0$. The function $f(\rho )$ is
negative for $\rho <r_{+}$ and positive for $\rho >r_{+}$, where $r_{+}$ is
the largest real root of $f(\rho )=0$. Indeed, $g_{\rho \rho }$ and $%
g_{\varphi \varphi }$ are $f(\rho )$ and $f^{-1}(\rho )g^{-1}(\rho )$, and
since $g(\rho )$ is positive both of $g_{\rho \rho }$ and $g_{\varphi
\varphi }$ become negative when $f(\rho )<0$ (which occurs for$\rho <r_{+}$%
). This leads to an apparent change of signature of the metric from $(n-1)+$
to $(n-2)+$ as one extends the spacetime to $\rho <r_{+}$, which is not
allowed. So, one cannot extend the spacetime to $\rho <r_{+}$. To get rid of
this incorrect extension, we introduce the new radial coordinate $r$ as
\[
r^{2}=\rho ^{2}-r_{+}^{2}\Rightarrow d\rho ^{2}=\frac{r^{2}}{r^{2}+r_{+}^{2}}%
dr^{2}.
\]
With this new coordinate, the metric (\ref{Met1}) is
\begin{eqnarray}
ds^{2} &=&-\frac{r^{2}+r_{+}^{2}}{l^{2}}R^{2}(r)dt^{2}+l^{2}f(r)d\varphi ^{2}
\nonumber \\
&&+\frac{r^{2}}{(r^{2}+r_{+}^{2})f(r)g(r)}dr^{2}+\frac{r^{2}+r_{+}^{2}}{l^{2}%
}R^{2}(r)dX^{2},  \label{metric2}
\end{eqnarray}
where the coordinates $r$ assumes the values $0\leq r<\infty $, and $f(r)$, $%
g(r)$, $R(r)$ and $\Phi (r)$ are now given as
\begin{eqnarray}
f(r) &=&\frac{r^{2}+r_{+}^{2}}{l^{2}}\Gamma ^{\gamma }-\frac{c^{n-2}}{%
(r^{2}+r_{+}^{2})^{(n-2)/2}}\Gamma ^{1-\gamma (n-2)},  \nonumber \\
g(r) &=&\Gamma ^{(n-3)\gamma },  \nonumber \\
R(r) &=&\Gamma ^{\gamma /2},  \nonumber \\
\Phi (r) &=&\frac{{(n-1)}{\alpha }}{{2}{(n-2+\alpha ^{2})}}\ln \Gamma , \nonumber\\
\Gamma &=& 1+ \frac{b^{n-2}}{(r^{2}+r_{+}^{2})^{(n-2)/2}} .
\label{Fr}
\end{eqnarray}

The function $f(r)$ is positive in the whole spacetime and is zero at $r=0$.
One can easily show that the Kretschmann scalar and all other curvature
invariants (such as Ricci scalar, Ricci square and so on) do not diverge in
the range $0\leq r<\infty $. However, the Taylor expansion of the metric in
the vicinity of $r=0$ is
\begin{equation}
ds^{2}=\frac{r_{+}^{2}}{l^{2}}\Gamma _{0}^{\gamma }[-dt^{2}+dX^{2}]+\frac{1}{%
r_{+}^{2}G_{0}}\left( dr^{2}+H_{0}(lr)^{2}d\varphi ^{2}\right) ,
\end{equation}
where
\begin{eqnarray}
\Gamma _{0} &=&\Gamma (r=0)=1+\left( \frac{b}{r_{+}}\right) ^{n-2}, \\
G_{0} &=&\frac{1}{2}\left[ \Gamma ^{(n-3)\gamma
}\frac{d^{2}f}{dr^{2}}\right] _{r=0}=\frac{1}{2l^{2}}\Gamma
_{0}^{(n-2)\gamma }[n-\frac{(\gamma+1) (n-2)}{(n-2)\gamma-1%
}(\frac{1}{\Gamma _{0}}-1)], \\\nonumber
H_{0} &=&\frac{1}{4}r_{+}^{2}\left[ \Gamma ^{(n-3)\gamma }(\frac{d^{2}f}{%
dr^{2}})^{2}\right] _{r=0}=\frac{r_{+}^{2}}{4l^{4}}\Gamma
_{0}^{(n-1)\gamma }[n-\frac{(\gamma+1) (n-2)}{(n-2)\gamma-1%
}(\frac{1}{\Gamma _{0}}-1)]^{2},
\end{eqnarray}
which shows that the spacetime has a conic geometry with a conical
singularity at $r=0$, since
\begin{equation}
\lim_{r\rightarrow 0}\frac{1}{r}\sqrt{\frac{g_{\varphi \varphi }}{g_{rr}}}%
\neq 1.  \label{limit}
\end{equation}
That is, as the radius $r$ tends to zero, the limit of the ratio
``circumference/radius'' is not $2\pi $, and therefore the spacetime has a
conical singularity at $r=0$. The canonical singularity can be removed if
one identifies the coordinate $\varphi $ with the period
\begin{equation}
{Period}_{\phi }=2\pi \left( \lim_{r\rightarrow 0}\frac{1}{r}\sqrt{\frac{%
g_{\varphi \varphi }}{g_{rr}}}\right) ^{-1}=2\pi (1-4\tau ),
\end{equation}
where $\tau $ is given by
\begin{mathletters}
\begin{equation}
\tau=\frac{1}{4}\left[ 1-\frac{2l(n-2+{\alpha }^{2}){\Gamma _{0}}^{1-\left(
n-1\right) \gamma /2}}{\left\{ n+[(n-2)^{2}-n{\alpha }^{2}](1-\Gamma
_{0})\right\} r_{+}}\right] .\label{Tau}
\end{equation}

This metric describes a spacetime that has a conical singularity at $r=0$
with a deficit angle $\delta =8\pi \tau $, which is proportional to the
brane tension at $r=0$ \cite{Rand}. The brane tension increases as $\alpha $
increases and approaches
\end{mathletters}
\begin{equation}
\tau =\frac{1}{4}\,\left[ 1-\frac{2l}{n{r_{+}}}\left[ 1+\left( {\frac{b}{%
r_{+}}}\right) ^{n-1}\right] ^{-1/(n-2)}\right]
\end{equation}
as $\alpha $ goes to infinity.

\section{Spinning Magnetic Branes\label{Spin}}

Now, we want to endow our spacetime solution with a global rotation. In order to add angular
momentum to the spacetime, one may perform the following rotation boost in
the $t-\varphi $ plane
\begin{equation}
t\mapsto \Xi t-a\varphi \ \ \ \ \ \ \ \ \ \ \varphi \mapsto \Xi \varphi -%
\frac{a}{l^{2}}t,  \label{tphi}
\end{equation}
where $a$ is the rotation parameter and $\Xi =1+a^{2}/l^{2}$. Performing the
transformation (\ref{tphi}) to the metric (\ref{Met1}) we obtain

\begin{equation}
ds^{2}=-\frac{\rho^{2}}{l^{2}}R^{2}(\rho)\left( \Xi dt-ad\varphi
\right) ^{2}+\frac{d\rho ^{2}}{\rho^{2}fg}+l^{2}f(\rho)\left( \frac{a%
}{l^{2}}dt-\Xi d\varphi \right) ^{2}+\frac{\rho^{2}}{l^{2}}R^2(\rho)%
dX^{2},  \label{Met2a}
\end{equation}
where the metric functions are the same as those given in Eq. (\ref{frho1}) and (\ref{frho2}).
The nonvanishing electromagnetic field components become
\begin{eqnarray}
F_{\varphi \rho}&=&\frac{ql\Xi e^{4\alpha \Phi /(n-1)}}{(\rho R)^{n-1}{g^{1/2}}}  \label{Ftr}\\
F_{t \rho}&=&-\frac{a}{l^{2}\Xi }F_{\varphi \rho} \label{Fttr}
\end{eqnarray}

The transformation (\ref{tphi}) generates a new metric, because it is not a
permitted global coordinate transformation. This transformation can be done
locally but not globally \cite{Stach}. Therefore, the metrics (\ref{metric2}%
) and (\ref{Met2a}) can be locally mapped into each other but not globally,
and so they are distinct. Indeed, the metric (\ref{Met2a}) provides an
example of gravitational fields which is locally static but globally
stationary. Again, this spacetime has no horizon and curvature singularity,
but it has a conical singularity at $r=0$.

We also study the rotating solutions with more rotation parameters. We now
generalize the above solution given in Eq. (\ref{metric2}) with $k\leq
\lbrack n/2]$ rotation parameters, where $[n/2]$ is the integer part of $n/2$%
. This generalized solution can be written as
\begin{eqnarray}
ds^{2} &=&-\frac{\rho^{2}}{l^{2}}R^{2}(\rho)\left( \Xi dt-{{\sum_{i=1}^{k}}%
}a_{i}d\varphi ^{i}\right) ^{2}+f(\rho)\left( \sqrt{\Xi ^{2}-1}dt-\frac{\Xi
}{\sqrt{\Xi ^{2}-1}}{{\sum_{i=1}^{k}}}a_{i}d\varphi ^{i}\right) ^{2}
\nonumber \\
&&+\frac{dr^{2}}{f(\rho)g(\rho)}+\frac{r^{2}}{l^{2}(\Xi ^{2}-1)}%
R^{2}(\rho){\sum_{i<j}^{k}}(a_{i}d\varphi _{j}-a_{j}d\varphi _{i})^{2}+%
\frac{r^{2}}{l^{2}}R^{2}(\rho)dX^{2},  \nonumber \\
\Xi ^{2} &=&1+\sum_{i=1}^{k}\frac{a_{i}^{2}}{l^{2}}.  \label{Met2b}
\end{eqnarray}
The nonvanishing components of the electromagnetic field tensor are
\begin{eqnarray}
&& F_{\varphi^i \rho}=\frac{ql\Xi a_i e^{4\alpha \Phi /(n-1)}}{(\Xi^2-1)^{1/2}(\rho R)^{n-1}{g^{1/2}}} ,\nonumber\\
&& F_{t\rho }=-\frac{(\Xi ^{2}-1)}{\Xi a_{i}}F_{\varphi ^{i}\rho }.
\end{eqnarray}

\section{Conserved Quantities\label{Cons}}

Although our solutions are in the presence of dilaton field, they are asymptotically AdS
in six and higher dimensions. Thus, one can calculate the conserved quantities through the use of the
counterterm method inspired by AdS/CFT correspondence for these asymptotically AdS
solutions. This method has been used by many authors on Einstein gravity \cite{Count}. Our solutions are asymptotically AdS in six and higher
dimensions, and therefore we apply the counterterm method to calculate the
conserved quantities for $n\geq 5$. For asymptotically AdS solutions with
flat boundary $\widehat{R}_{abcd}(h)=0$, the finite energy-momentum tensor
is
\begin{equation}
T_{ab}=\frac{1}{8\pi }\{Kh_{ab}-K_{ab}-\frac{n-1}{l}h_{ab}\},
\label{Stres}
\end{equation}
To compute the conserved charges of the spacetime, we choose a spacelike
surface $\mathcal{B}$ in $\partial \mathcal{M}$ with metric $\sigma _{ij}$
and write the boundary metric in Arnowitt-Deser-Misner form:
\begin{equation}
h_{ab}dx^{a}dx^{a}=-N^{2}dt^{2}+\sigma _{ij}\left( d\varphi
^{i}+V^{i}dt\right) \left( d\varphi ^{j}+V^{j}dt\right) ,
\end{equation}
where the coordinates $\varphi ^{i}$ are the angular variables parametrizing
the hypersurface of constant $r$ around the origin and $N$ and $V^{i}$ are
the lapse and shift functions, respectively. When there is a Killing vector
field $\mathcal{\xi }$ on the boundary, then the quasilocal conserved
quantities associated with the stress tensors of Eq. (\ref{Stres}) can be
written as
\begin{equation}
\mathcal{Q}(\mathcal{\xi )}=\int_{\mathcal{B}}d^{n-1}\varphi \sqrt{\sigma }%
T_{ab}n^{a}\mathcal{\xi }^{b},  \label{charge}
\end{equation}
where $\sigma $ is the determinant of the metric $\sigma _{ij}$ and $n^{a}$
is the timelike unit normal vector to the boundary $\mathcal{B}$\textbf{.}
In the context of counterterm method, the limit in which the boundary $%
\mathcal{B}$ becomes infinite ($\mathcal{B}_{\infty }$) is taken, and the
counterterm prescription ensures that the action and conserved charges are
finite. For our case of horizonless rotating spacetimes, the first Killing
vector is $\xi =\partial /\partial t$, and therefore its associated
conserved charge of the brane is the mass per unit volume calculated as
\begin{equation}
M=\int_{\mathcal{B}_{\infty }}d^{n-1}\varphi \sqrt{\sigma }T_{ab}n^{a}\xi ^{b}=\frac{1}{16\pi}[n({\Xi}^2-1)+1]\left(\frac{c}{l}\right)^{n-2}.
\label{Mass}
\end{equation}
The second class of conserved quantities is the angular momentum per unit
volume associated with the rotational Killing vectors $\zeta _{i}=\partial
/\partial \phi ^{i}$, which may be calculated as
\begin{equation}
J_{i}=\int_{\mathcal{B}_{\infty }}d^{n-1}\varphi \sqrt{\sigma }T_{ab}n^{a}\zeta
_{i}^{b}=-\frac{\Xi }{8\pi }\left( \frac{c}{l}\right) ^{n-2}a_{i}.
\label{Ang}
\end{equation}
For $a_{i}=0$ ($\Xi =1$), the angular momentum per unit volume vanishes, and
therefore $a_{i}$'s are the rotational parameters of the spacetime.

Next, we calculate the electric charge of the solutions. To determine the
electric field, we should consider the projections of the electromagnetic
field tensor on special hypersurfaces. The normal to such hypersurfaces for
the spacetimes with a longitudinal magnetic field is
\[
u^{0}=\frac{1}{N},{\ \ }u^{r}=0,{\ \ }u^{i}=-\frac{N^{i}}{N},
\]
and the electric field is $E^{\mu }=g^{\mu \rho }F_{\rho \nu }u^{\nu }$.
Then the electric charge per unit volume can be found by calculating the
flux of the electromagnetic field at infinity, yielding
\begin{equation}
Q=\frac{1}{4\pi}\frac{q\sqrt{\Xi ^{2}-1}}{l^{n-2}}. \label{elecch}
\end{equation}
Note that the electric charge is proportional to the rotation parameter and
is zero for the case of static solutions. The electric charge (\ref{elecch})
and the conserved quantities given by Eqs. (\ref{Mass}) and (\ref{Ang}) show
that the metrics (\ref{Met1}) and (\ref{Met2b}) cannot be mapped into each
other globally.

\section{Closing Remarks}

The maximally symmetric spacetimes are not the solutions of field equations
of dilaton gravity in the presence of one or two Liouville-type potentials,
but an appropriate combination of three Liouville-type potentials can
support maximally symmetric spacetimes in six and higher-dimensions. In this
paper, we found a new class of magnetic solutions in Einstein-Maxwell-dilaton
gravity in the presence of three Liouville-type potentials, which are
asymptotically AdS in six and higher-dimensions, and investigated the
effects of the dilaton field on the properties of the spacetime. This class
of solutions yields an $(n+1)$-dimensional spacetime with a longitudinal
nonlinear magnetic field [the only nonzero component of the electromagnetic
field is $F_{r\varphi }$] generated by a static magnetic brane. We found
that these solutions have no curvature singularity and no horizons, but have
conic singularity at $r=0$ with a deficit angle which is sensitive to the
dilaton field. We found that, as the coupling constant $\alpha $ becomes
larger, the brane tension increases and approaches to a constant as $\alpha $
goes to infinity. Also, it is worth to mention that one can have a curved
spacetime without horizon or any kind of singularities (curvature, or conic)
by fine tuning of the parameters of the spacetime such that the brane
tension given in Eq. (\ref{Tau}) vanishes.

We found that for the case of static spacetime, the electric field vanishes,
and therefore the brane has no net electric charge. Next, we endow the
spacetime with rotation. For the spinning brane, when the rotation
parameters are nonzero, the brane has a net electric charge density which is
proportional to the magnitude of the rotation parameters given by $(\Xi
^{2}-1)^{1/2}$. We also applied the counterterm method inspired by AdS/CFT
correspondence in order to calculate the conserved quantities of the
spacetime in six and higher dimensions and found that these conserved
quantities do not depend on $\alpha $. This can be understand easily, since
at the boundary at infinity the dilaton field vanishes ($\Phi (\infty )=0$).

\begin{acknowledgements}
This work has been supported financially by Research Institute for
Astronomy and Astrophysics of Maragha.
\end{acknowledgements}

\end{document}